\documentclass[11pt]{article}
\usepackage{amssymb,amsmath,amsfonts}
\usepackage{graphicx}
\usepackage{graphics}
\usepackage{eepic,epsfig}
\usepackage{verbatim}
\usepackage{color}
\usepackage{hyperref}
\usepackage{cite}

1.60cm \makeatletter \@addtoreset{equation}{section}

\makeatother

\begin{document}
\title{Scalar self-interaction in the spacetime of a cosmic dispiration}
\author{K. E. L. de Farias$^1$\thanks{E-mail: klecio.lima@academico.ufpb.br}\ , E. A. F. Bragan\c{c}a$^2$\thanks{E-mail: eduardo.braganca@uemasul.edu.br} \ and H. F. Santana Mota$^1$\thanks{E-mail: hmota@fisica.ufpb.br} \,
\\
\textit{$^{1}$Departamento de F\'{\i}sica, Universidade Federal da Para\'{\i}ba 58.059-970,}\\
\textit{Caixa Postal 5.008,Jo\~{a}o Pessoa, PB, Brazil}\\
\textit{$^{2}$Centro de Ci\^{e}ncias Agr\'{a}rias, Naturais e Letras - CCANL,}\\
\textit{Universidade Estadual da Regi\~{a}o Tocantina do Maranh\~{a}o},\\
\textit{Avenida Brejo do Pinto S/N, 65975-000, Estreito, MA, Brazil}
\vspace{0.3cm}\\}
\maketitle
%
\begin{abstract}
In the present paper we investigate the classical self-interaction associated with a charged scalar point particle placed at rest in the gravitational field of a linear topological defect known in literature as cosmic dispiration, a combination of a cosmic string with a screw dislocation. We found exact expressions for the Green's function and, consequently, for the self-energy by considering both massive and massless scalar point particles. We present numerical graphs and analyze the nature of the self-interactions in the cosmic dispiration spacetime itself and in the cases of pure screw dislocation and cosmic string spacetimes. 
\end{abstract}

\section{Introduction}
\label{Int}

The physical phenomenon of a point particle interacting with its own classical field traces back at least to the first half of the twentieth century, when attempts were made to take into account, in the equation of motion, the self-interaction of an electrically charged point particle \cite{AB, Lorentz, dirac1938classical}. The electromagnetic field created when the particle is set in motion diverges at its position, such that a renormalization procedure is adopted in order to obtain an equation of motion whose solutions are finite \cite{AB, Lorentz, dirac1938classical}. Since then, several works have been done in attempting to address some issues associated with the original proposal, as well as to generalize it to curved spacetimes (see \cite{Khusnutdinov:2020mkz, Pfenning:2000zf, Poisson:2011nh} for a review). 

The investigation of the physics involving self-interactions has also been extended to the case of a self-gravitational force on a massive point particle, and to the case of a self-force on a charged scalar particle \cite{Khusnutdinov:2020mkz, Quinn:2000wa, Pfenning:2000zf, Poisson:2011nh}. In particular, in Ref. \cite{Quinn:2000wa}, a rigorous axiomatic approach has been proposed to study the self-interaction arising from the motion of a charged scalar point particle in curved spacetime. In this framework, it was shown that an action principle can be formulated, providing the correct equation of motion, as long as the particle's mass varies with proper time. This is a consequence of a contribution to the self-interaction in curved spacetime which is non-local in time, that is, the past history of the particle plays an important role in its motion. 

Self-interactions in curved spacetime emerge even when the point particle is placed at rest, in which case its state of motion does not depend on time. In this direction, several works have been conducted considering different spacetimes \cite{Khusnutdinov:2020mkz, smith1980force, linet1986force, Linet:1986db, Azevedo:2000nx, escidoc:153364, DeLorenci:2001sn,  BezerradeMello:2006hz, Barbosa:2010sx}. In particular, self-forces on charged point particles induced by the gravitational field of a conical defect, like the one of a cosmic string or disclination, have been considered in Refs. \cite{linet1986force, Linet:1986db, Azevedo:2000nx, escidoc:153364}. In this case, as the spacetime of a conical defect is locally flat, there is no direct gravitational field acting on the particle but, instead, only a self-force is produced. This also happens when the particle is set at rest in the spacetime of a cosmic dispiration, the combination of a cosmic string with a screw dislocation, as shown in \cite{DeLorenci:2001sn}, where the electrostatic and gravitational self-forces were investigated.

Cosmic strings are particularly interesting linear topological defects that are expected to be formed in the early universe, and may have left gravitational, cosmological and astrophysical imprints \cite{Copeland:2011dx, Hindmarsh:2011qj, hindmarsh, VS, Mota:2014uka}. No less interesting is the spacetime of a screw dislocation, which is also a linear topological defect, and exists in the context of theories of solid and crystal continuum media, much studied in condensate matter systems \cite{Puntigam:1996vy}. The combination of a cosmic string with a screw dislocation to form the spacetime of a cosmic dispiration is found in the framework of the Einstein-Cartan theory of gravity and presents a spacelike helical structure, hence, it can be known as twisted cosmic string \cite{Letelier:1995ze, Galtsov:1993ne}. This structure is also obtained when the spacetime describing a (2+1) spinning particle is lifted to (3+1)-dimensions in such way that the boost covariance is preserved \cite{Galtsov:1993ne}. Motivated by the richness of the physics involved in the structure of these topological defects, in this paper, we wish to investigate the self-interaction induced by placing a charged scalar point particle at rest in the spacetime of a cosmic dispiration \cite{mota2018scalar}. This has not been considered previously in literature, to the best of our knowledge.

This paper is organized as follows. In Sec.\ref{background} we present the line element describing the cosmic dispiration spacetime, the action which gives the equation of motion sourced by a scalar point particle and the expression to calculate the self-energy. In Sec.\ref{sec3} we develop our calculations by solving the equation of motion and, consequently, the Green's function that makes possible to obtain the self-energy is found. Finally, in Sec.\ref{Con} we present our conclusions. Throughout the paper we use natural units $G=c=1$.
%
\section{Background geometry and equation of motion}
\label{background}
%
Firstly, we want to lay out the physical configuration at which we wish to investigate the scalar self-interaction. For this purpose, let us consider the line element that describes an idealized $(3+1)-$dimensional cosmic dispiration spacetime, composed by a cosmic string and a screw dislocation. In cylindrical coordinates, the line element of such a spacetime is given by \cite{mota2018scalar, DeLorenci:2001sn}
\begin{equation}
ds^2=g_{\mu\nu}dx^\mu dx^\nu=dt^2-dr^2-r^2d\phi^2-(dz+\kappa d\phi)^2,
\label{line1}
\end{equation}
where $(r,\phi,z)$ are the cylindrical coordinates
which take on the values $r\geq0$, $0\leq\phi\leq \phi_0=2\pi/\nu$ and $-\infty<(t,z)<+\infty$. The presence of a cosmic string
is codified by the parameter $\nu\geq1$, which is given by $\nu^{-1}=1-4G\mu_0$, with $G$ being the Newton's gravitational
constant and $\mu_0$ the linear mass density of the string \cite{hindmarsh, VS}. In condensate matter systems, however, the parameter $\nu$ characterizes a disclination and can assume any values such that $\nu\geq0$ \cite{Katanaev:1992kh}. The parameter $\kappa$ is a constant characterizing the screw dislocation, known as Burgers vector in theories of continuum media \cite{Puntigam:1996vy}, but associated with $\frac{2GJ_z}{\pi}$ in the framework of Einstein-Cartan theory of gravity, with $J^z$ being the $z$-component of a vector tangent to the string world sheet  \cite{Letelier:1995ze, Galtsov:1993ne}. The spacetime described by the line element \eqref{line1} is locally flat and has the nontrivial identification condition $(t, r, \phi, z)\rightarrow (t, r, \phi + 2\pi/\nu, z)$ \cite{mota2018scalar, DeLorenci:2001sn}.

The line element \eqref{line1} can also be written in the following form \cite{mota2018scalar}:
\begin{equation}
ds^2=dt^2-dr^2-r^2d\phi^2-dZ^2,
\label{spacesection}
\end{equation}
where we have defined the new spatial coordinate $Z=z + \kappa\phi$, leading to the new spacetime identification condition $(t, r, \phi, Z)\rightarrow (t, r, \phi + 2\pi/\nu, Z + 2\pi\kappa/\nu)$ \cite{mota2018scalar, DeLorenci:2001sn}. Note that the line element written in the form given by \eqref{spacesection} shows that the cosmic dispiration spacetime is locally flat. In our study, we shall use the line element describing the (3+1)-dimensional cosmic dispiration spacetime as given by \eqref{spacesection} \cite{mota2018scalar}.

It has been shown that the mass of a point scalar charge, once we take its self-interaction into consideration in curved spacetime, is no longer a constant of motion but instead vary with proper time \cite{Quinn:2000wa}. Contributions both from the curvature of the spacetime and past history of the particle's dynamics inevitably take place. This is a necessary feature in order to obtain the correct equations of motion, for the particles' dynamics, from a variational principle formulation. Nevertheless, here, we are interested in considering the self-interaction energy of a point scalar charge at rest, in which case the mass is constant and the only contribution arising is the one from the topology of the spacetime, namely, the cosmic dispiration spacetime. The self-interaction, then, arises solely due to the massive scalar field, $\varphi$, being coupled to the charge density, $\rho$, of the point scalar particle. In curved spacetime, the action describing this system in $(3+1)$-dimensional cosmic dispiration spacetime is written as
%
\begin{equation}
S=\frac{1}{2}\int d^{4}x\sqrt{|g|} \left[g^{\mu\nu}\nabla_\mu\varphi\nabla_\nu\varphi-(\xi R +m^2)\varphi^2\right]
+\int d^{4}x\sqrt{|g|}\rho\varphi,
\label{action}
\end{equation}
where $g=\text{det}(g_{\mu\nu})$, $\xi$ is the constant coupling to gravity, $R$ is the Ricci scalar and $\nabla_{\mu}$ is the covariant derivative. Note that we are using the following notation: $x=(t, {\bf r})=(t,r,\phi, Z)$. Thereby, the first term on the right-hand side of Eq. \eqref{action} is the Klein-Gordon action considering an arbitrary curvature
coupling $\xi$, whereas the second term presents the interaction term.

The standard procedure to obtain the field equation is to vary the action \eqref{action} with respect to $\varphi(x)$. As we are considering a scalar point particle at rest, there will be no time-dependence on the scalar field, and the resulting equation of motion is obtained as 
\begin{equation}
\left(-\nabla^2+m^2\right)\varphi=\rho,
\label{SKG}
\end{equation}
where $\nabla^2$ is the Laplace operator in three spatial dimensions and the scalar charge density, for a point particle, is given by 
\begin{equation}
\rho({\bf r}) = q\frac{\delta^3({\bf r} - {\bf r}_p)}{\sqrt{|g|}},
\label{CD}
\end{equation}
with $q$ being the scalar charge and ${\bf r}_p$ the position of the particle. Note that we have set $\xi=0$, characterizing a minimally coupled scalar field. We have chosen to consider this case in order to avoid the complications associated with the infinite curvature contribution, at $r=0$, due to the presence of the cosmic string. Such a contribution to the scalar field, $\varphi(x)$, is a delta-like contribution and has been worked out in Refs. \cite{Jackiw:1991je, Spinally:2000ii}.

The classical self-energy of a scalar point particle is formally obtained by considering the 00-component of the energy-momentum tensor. An expression for the latter, on the other hand, is provided by the variation of the action \eqref{action} with respect to the metric. This gives
\begin{equation}
T_{\mu\nu}=-g_{\mu\nu}\rho\varphi+\nabla_\mu\varphi \nabla_\nu\varphi-\frac{1}{2}g_{\mu\nu}
\left(g^{\alpha\beta}\nabla_\alpha \varphi \nabla_\beta \varphi-m^2\varphi^2\right)
-\xi\left(G_{\mu\nu}\varphi^2-g_{\mu\nu}\Box\varphi^2+\nabla_\mu\nabla_\nu\varphi^2\right),
\label{tensor}
\end{equation}
where $G_{\mu\nu}$ is the Einstein tensor and $\Box$ is the d'Alembert operator. Thus, by integrating $T^0_0$ over the spatial volume we obtain the following result:
\begin{equation}
U=-\frac{1}{2}\int d^3{\bf r} \, \sqrt{|g|} \, \rho({\bf r})\varphi({\bf r}),
\label{SE}
\end{equation}
which is the formal self-energy expression to be used in our calculations. Note that this expression depends on the solution, $\varphi({\bf r})$, of the equation \eqref{SKG} which in turn is given in terms of the Green's function $G({\bf r,r'})$ as
\begin{equation}
\varphi({\bf r}) = \int d^3{\bf r}\sqrt{|g|} G({\bf r,r'})\rho({\bf r'}).
\end{equation}
In fact, this is a solution of \eqref{SKG} as long as the Green's function obeys the equation
\begin{equation}
(-\nabla^2+m^2)G({\bf r,r'})=\frac{\delta^3({\bf r}-{\bf r'})}{\sqrt{|g|}}.
\label{green0}
\end{equation}
Hence, the self-energy expression \eqref{SE} becomes 
\begin{eqnarray}
U=-\frac{q^2}{2}\, G({\bf r}_p,{\bf r}_p),
\label{Uelec2}
\end{eqnarray}
which depends almost exclusively on the Green's function calculated at the particle's position. Therefore, once we find the solution to \eqref{green0} we will be able to obtain the scalar point particle self-energy \eqref{Uelec2}. Note that the self-force is obtained as ${\bf F}=-\nabla U$. In what follows we will preferable work with the self-energy instead of the self-force itself, which in the case of a massive scalar particle provides lengthy expressions. This does not prevent us to understand the behaviour of the particle under the action of its own generated force.

\section{Green's function and self-energy}
\label{sec3}
%
We want to start the present section by calculating the Green's function $G({\bf r,r'})$. As it obeys Eq. \eqref{green0}, we can expand it in terms of the eigenfunctions, $\Phi_{\sigma}$, of the eigenvalue equation constructed with the same operator, $(-\nabla^2+m^2)$, that acts in the Green's function \cite{butkov1968mathematical}. That is,
\begin{eqnarray}
(-\nabla^2+m^2)\Phi_{\sigma}({\bf r}) = \lambda_{\sigma}\Phi_{\sigma}({\bf r}),
\label{EVE}
\end{eqnarray}
where $\lambda_{\sigma}$ are the eigenvalues of $\Phi_{\sigma}({\bf r}) $ and 
\begin{equation}
\nabla^2=\frac{1}{r}\frac{\partial}{\partial r}\left(r\frac{\partial}{\partial r}\right)+\frac{1}{r^2}\left(\frac{\partial}{\partial\phi}+\kappa\frac{\partial}{\partial z}\right)^2+\frac{\partial^2}{\partial z^2},
\label{NOp}
\end{equation}
is given in cylindrical coordinates, taken into consideration the spatial section of \eqref{line1}, the line element describing the cosmic dispiration spacetime. Consequently, the Green's function can be calculated by means of the expansion
\begin{eqnarray}
G({\bf r,r'}) =\sum_{\sigma}\frac{\Phi_{\sigma}({\bf r})\Phi_{\sigma}({\bf r'})}{ \lambda_{\sigma}}. 
\label{GFexp}
\end{eqnarray}

The complete set of normalized solutions of Eq. \eqref{EVE}, with \eqref{NOp}, has been obtained, for instance, in Ref. \cite{mota2018scalar}. In three spacial dimensions, it is written as
\begin{equation}
\Phi_{\sigma}({\bf r})=\left[\frac{\nu\eta}{(2\pi)^2}\right]^{\frac{1}{2}}J_{|n\nu-\kappa k_z|}(\eta r)e^{i(n\nu-\kappa k_z)\phi+ik_z Z},
\label{eigenfunction2}
\end{equation}
where $J_\alpha(x)$ is the Bessel function of the first kind. For our particular physical system, it is clear that $\sigma$ is the set of the eigenfunctions modes and $\lambda_{\sigma}=m^2+\eta^2+k_z^2$ are the eigenvalues of $\Phi_{\sigma}({\bf r})$, satisfying the eigenvalue equation \eqref{EVE}. Note that the summation symbol in Eq. \eqref{GFexp}, in this case, stands for
\begin{equation}
\sum_{\sigma}=\sum_{n=-\infty}^{\infty}\int_{-\infty}^{\infty} dk_z\int_0^{\infty} d\eta.
\label{SumS}
\end{equation}

The Green's function \eqref{GFexp}, by making use of Eqs. \eqref{eigenfunction2} and \eqref{SumS}, now can be calculated. Thus, by using the useful identities (A5) and (B4) of Ref. \cite{Mota:2016mhe}, the Green's function can be put in the form 
\begin{eqnarray}
G({\bf r,r'}) &=&\frac{\nu}{2(2\pi)^{2}}\int_0^{\infty}\frac{dw}{w}\, e^{-\frac{m^2rr'}{2w}-\frac{w(r^2 + r'^2)}{2rr'}} \,
\mathcal{I}(w,\kappa,\nu),
\label{green3}
\end{eqnarray}
where we have made the change of variable $w=\frac{rr'}{2t}$ and 
\begin{equation}
\mathcal{I}(w,\kappa,\nu)=\int_{-\infty}^{+\infty}\frac{\nu dh}{\kappa}e^{-\frac{\nu^2h^2rr'}{2w\kappa^2}+\frac{i\nu h\Delta Z}{\kappa}}
\sum_{n=-\infty}^{\infty}e^{i\nu(n-h)\Delta \phi}I_{\nu|n-h|}(w),
\label{representation1}
\end{equation}
with $\Delta\phi=\phi-\phi'$, $\Delta Z=Z-Z'$, $h=\frac{\kappa k_z}{\nu}$ and $I_\alpha(x)$ is the modified Bessel function of the first kind. In Ref. \cite{mota2018scalar} the authors have developed a useful expression for the function $\mathcal{I}(w,\kappa,\nu)$, given by Eq. (A20) in \cite{mota2018scalar}. This makes possible to perform the integral in $w$ present in \eqref{green3} so that the final form for the Green's function is found to be
\begin{eqnarray}
G({\bf r,r'}) =\frac{m}{(2\pi)^{\frac{3}{2}}}\left[\sum_l  f_{\frac{1}{2}}(m\sigma_l)-\frac{\nu}{\pi^2}
\sum_{n=-\infty}^{\infty}\int_0^\infty dy \, f_{\frac{1}{2}}(m\sigma_{y,n})M_{n,q}(\Delta\phi,y)\right],
\label{green4}
\end{eqnarray}
where we have defined the function $f_\mu(x)$ as
\begin{equation}
f_\mu(x)=\frac{K_\mu(x)}{x^\mu},
\end{equation}
with $K_\mu(x)$ being the  Macdonald function and
\begin{eqnarray}
\sigma_l^2&=&r^2 + r'^2-2rr'\cos(2\pi l/q-\Delta\phi) + \left(\Delta Z - \frac{2\pi\kappa l}{\nu}\right)^2\nonumber,
\\
\sigma_{y,n}^2&=&r^2 + r'^2+2rr'\cosh(y)+\left(\Delta Z + \frac{2\pi\kappa n}{\nu}\right)^2.
\label{Gd}
\end{eqnarray}
Moreover, the function $M_{n,q}(\Delta\phi,y)$ is expressed as \cite{mota2018scalar}
\begin{equation}
M_{n,\nu}(\Delta\phi,y)=\frac{1}{2}\left\{\frac{\left(n+\frac{\nu}{2}+{\frac{\Delta\phi}{\phi_0}}\right)}
{\left(n+\frac{\nu}{2}+\frac{\Delta\phi}{\phi_0}\right)^2+\left(\frac{y}{\phi_0}\right)^2}
-\frac{\left(n-\frac{\nu}{2}+{\frac{\Delta\phi}{\phi_0}}\right)}
{\left(n-\frac{\nu}{2}+\frac{\Delta\phi}{\phi_0}\right)^2+\left(\frac{y}{\phi_0}\right)^2}\right\},
\label{Mfunction}
\end{equation}
where $\phi_0=\frac{2\pi}{\nu}$ and the discrete sum index $l$ in Eq. \eqref{green4} must obey the restriction 
\begin{equation}
-\frac{\nu}{2}+\frac{\Delta\phi}{\phi_0}\leq l \leq \frac{\nu}{2}+\frac{\Delta\phi}{\phi_0}.
\end{equation}

One should point out that the Green's function obtained in Eq. \eqref{green4} diverges in the coincidence limit ${\bf r'}\rightarrow {\bf r}$, where the self-energy is calculated. In fact, the divergent contribution comes from the $l=0$ term in the first term on the r.h.s of \eqref{green4} and needs to be subtracted. This divergent $l=0$ term is given by 
\begin{eqnarray}
G_0({\bf r,r'}) =\frac{m}{(2\pi)^{\frac{3}{2}}}f_{\frac{1}{2}}(m\sigma_0),
\label{divterm}
\end{eqnarray}
where $\sigma_0$ is the geodesic distance between two points which is given by \eqref{Gd}, for $l=0$. Note that, by using the asymptotic limit of $K_{\mu}(x)$ for small arguments \cite{abramowitz},  the massless scalar field limit ($m\rightarrow 0$) of Eq. \eqref{divterm} is given by
\begin{eqnarray}
G_0({\bf r,r'}) =\frac{1}{4\pi\sigma_0},
\label{divtermass}
\end{eqnarray}
which exactly coincides with the leading term in the the DeWitt-Schwinger expansion for the three-dimensional Green's function \cite{birrell1984quantum}, as it should be since it is the term to be subtracted out.

The formal procedure to calculate the self-energy \eqref{Uelec2} in curved spacetime is, thus, written as
\begin{eqnarray}
U_{\text{ren}}=-\frac{q^2}{2}\lim_{{\bf r'}\rightarrow {\bf r}}[G({\bf r},{\bf r'}) - G_0({\bf r},{\bf r'})]. 
\label{Uelec2ren}
\end{eqnarray}
\begin{figure}[h]
	\centering
	{\includegraphics[width=1.0\textwidth]{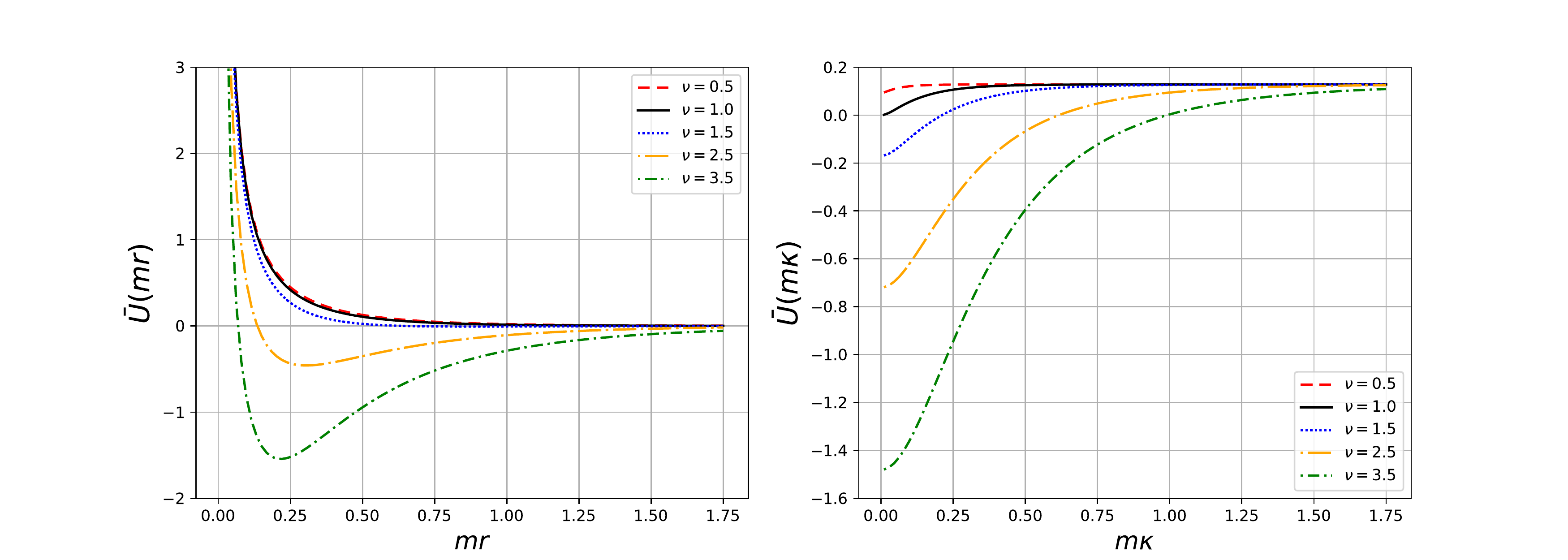}}
	\caption{Behaviour of the dimensionless massive self-energy, $\bar{U}=\frac{2(2\pi)^{\frac{3}{2}}U_{\text{ren}}}{mq^2}$, in Eq. \eqref{green4ren}, as a function of $mr$ in the plot on the left and as a function of $m\kappa$ in the plot on the right. We have also taken $m\kappa = 0.25$ for the plot on the left and $mr=0.5$ for the plot on the right.}
	\label{f1}
\end{figure}
By taking into consideration the expressions for the Green's function \eqref{green4} and \eqref{divterm} we find that the renormalized and finite self-energy is obtained as 
\begin{eqnarray}
U_{\text{ren}}&=&-\frac{mq^2}{2(2\pi)^{\frac{3}{2}}}\left[2\sideset{}{'}\sum_{l=1}^{[\nu/2]}  f_{\frac{1}{2}}\left(\sqrt{(2mrs_l)^2 + (m\bar{\kappa}l)^2}\right)\right.\nonumber\\
&&\left. -\frac{\nu}{\pi^2}
\sum_{n=-\infty}^{\infty}\int_0^\infty dy \, f_{\frac{1}{2}}\left(\sqrt{(2mrs_y)^2 + (m\bar{\kappa}n)^2}\right)M_{n,\nu}(y)\right],
\label{green4ren}
\end{eqnarray}
where $\bar{\kappa}=\frac{2\pi\kappa}{\nu}$, $s_l=\sin(\pi l/\nu)$, $s_y=\cosh(y/2)$ and the function \eqref{Mfunction} in the coincidence limit reduces to 
\begin{equation}
M_{n,\nu}(y)=\frac{\left(n+\frac{\nu}{2}\right)}
{\left(n+\frac{\nu}{2}\right)^2+\left(\frac{y}{\phi_0}\right)^2}. 
\label{Mfunction2}
\end{equation}
Note that in Eq. \eqref{green4ren} the notation $[\nu/2]$ stands for the integer part of $\nu/2$ and the prime symbol present in the sign of summation in $l$ means that in the case $\nu$ is an integer number the sum should be replaced with \cite{mota2018scalar, Mota:2016mhe}
\begin{equation}
\sideset{}{'}\sum_{l=1}^{[\nu/2]}\rightarrow \frac{1}{2}\sum_{l=1}^{\nu-1}.
\label{changesum}
\end{equation}
Therefore, the renormalized self-energy in the cosmic dispiration spacetime \eqref{line1} has a closed and exact form, as we can see from Eq. \eqref{green4ren}. It depends on the cosmic string and screw dislocation parameters $\nu$ and $\kappa$, respectively. Note that although $K_{\frac{1}{2}}(x)=\sqrt{\frac{\pi}{2x}}e^{-x}$ (see \cite{abramowitz}) we have chosen to keep the notation $f_{\frac{1}{2}}(x)$ which is somewhat shorter. In any case, this shows that the massive scalar self-energy \eqref{green4ren} is exponentially suppressed for large arguments. The massive scalar self-energy \eqref{green4ren} is plotted in Fig.\ref{f1}, in terms of $m\kappa$ on the right and in terms of $mr$ on the left. These plots indicate that depending on the values of the cosmic string and screw dislocation parameters, the self-interaction can be repulsive or attractive. For instance, the curve for $\nu=3.5$ on the left plot suggests that the self-force is repulsive for small values of $mr$, vanishes at around $mr\simeq 0.24$ and, then, becomes attractive, going to zero as $mr\rightarrow\infty$. The curves for $\nu\leq 1.5$, on the other hand, indicate a repulsive self-force. 

\begin{figure}[h]
	\centering
	{\includegraphics[width=0.7\textwidth]{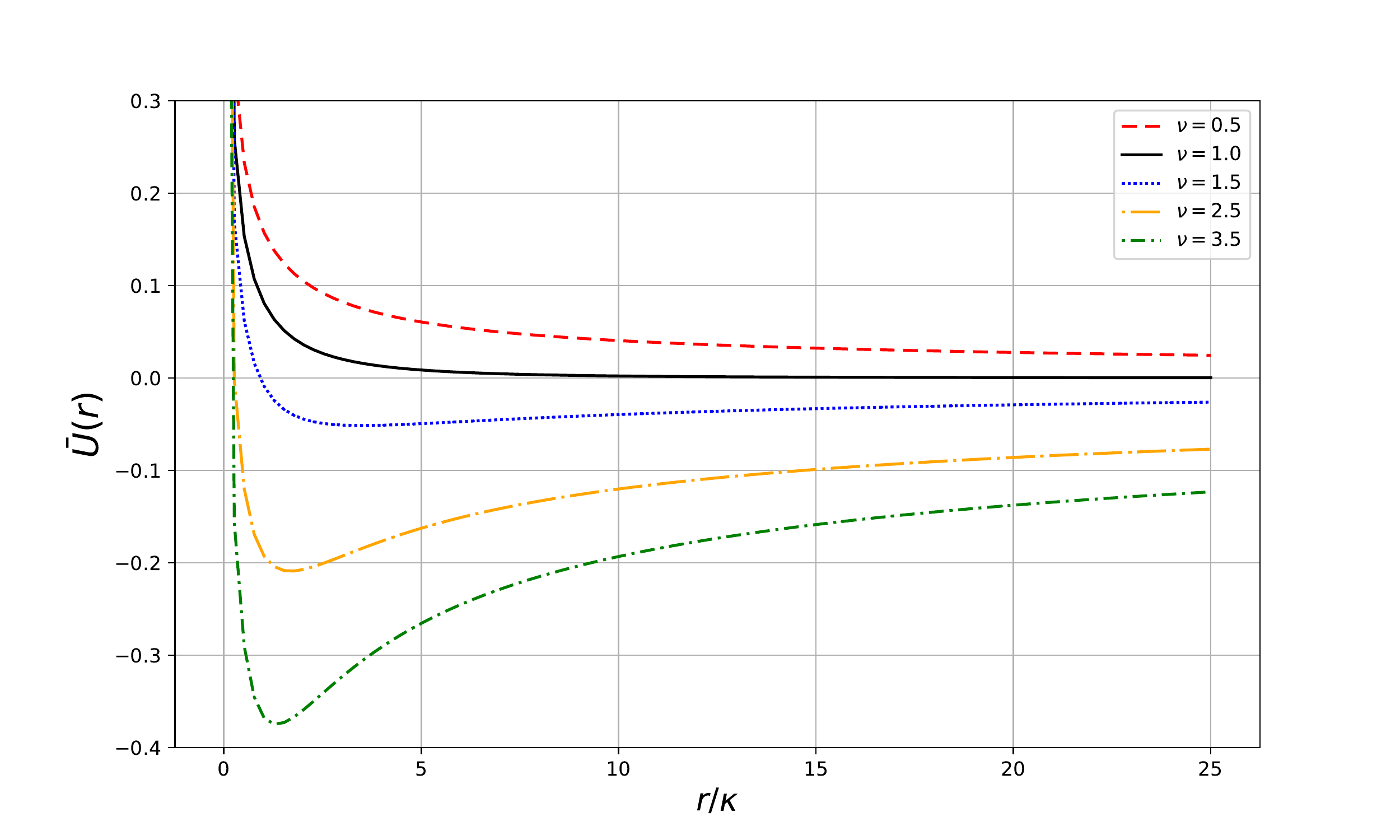}}
	\caption{Behaviour of the dimensionless massless self-energy, $\bar{U}=\frac{4\pi\kappa U_{\text{ren}}}{q^2}$, in Eq. \eqref{green4renmassless}, as a function of $r/\kappa$.}
	\label{f2}
\end{figure}

The massless scalar field self-energy is obtained in the limit $m\rightarrow 0$ of Eq. \eqref{green4ren}. This gives  
\begin{eqnarray}
U_{\text{ren}}&=&-\frac{q^2}{4\pi}\left[\sideset{}{'}\sum_{l=1}^{[\nu/2]} \left[4r^2s_l^2 + (\bar{\kappa}l)^2\right]^{-\frac{1}{2}}\right.\nonumber\\
&&\left. -\frac{\nu}{2\pi^2}
\sum_{n=-\infty}^{\infty}\int_0^\infty dy \, \left[4r^2s_y^2 + (\bar{\kappa}n)^2\right]^{-\frac{1}{2}}M_{n,\nu}(y)\right],
\label{green4renmassless}
\end{eqnarray}
where we have made use of the asymptotic form of the Bessel function $K_{\alpha}(x)$, for small arguments \cite{abramowitz}. This expression for the self-energy of a massless scalar particle is plotted in Fig.\ref{f2} in terms of $\frac{r}{\kappa}$. It indicates a repulsive self-force for the values of the cosmic string parameter $\nu\leq 1$, and can be repulsive,  attractive or even vanish for $\nu>1$, similarly to the massive case. One should point out that in both expressions for the massive \eqref{green4ren} and massless \eqref{green4renmassless} scalar fields self-energy there will be no contribution from the first term on the r.h.s if the cosmic string parameter is assumed to be $\nu<2$ \cite{mota2018scalar, Mota:2016mhe}. Hence, the only contribution to the self-energies comes from the second term on the r.h.s. In the particular case in which there is only a screw dislocation, i.e., $\nu=1$, the only contribution to the self-energies also comes from the second term on the r.h.s of Eqs. \eqref{green4ren} and \eqref{green4renmassless}, with $\nu=1$. By the left plot of Fig.\ref{f1}, we can infer that the self-force of a point scalar particle in the screw dislocation spacetime is repulsive for the particular value of $m\kappa=0.25$. In the massless case, we can also infer that from the plot in Fig.\ref{f2}.

Let us now analyse the asymptotic limits when the particle is far away ($mr\gg 1$) and near ($mr\ll 1$) the cosmic dispiration, keeping $m\bar{\kappa}$ fixed. The resulting expressions will depend on what happens with $\sqrt{(2mrs)^2 + (m\bar{\kappa}j)^2}$ in the argument of the function $f_{\mu}(x)$ in Eq. \eqref{green4ren}. Near the cosmic dispiration, for instance, $m\bar{\kappa}j$ dominates when $j=l$. When $j=n$, on the other hand, the dominant term near the defect is the one for $n=0$, resulting in $2mr\cosh(y/2)$ in the argument of $f_{\mu}(x)$. The self-energy near the cosmic dispiration, hence, takes the form   
\begin{eqnarray}
U_{\text{ren}}\simeq -\frac{q^2}{4\pi}\left[\frac{\nu}{2\pi\kappa}\sideset{}{'}\sum_{l=1}^{[\nu/2]}\frac{e^{-\frac{2\pi\kappa ml}{\nu}}}{l}-\frac{\nu}{4\pi^2r}\int_0^{\infty}dy\frac{e^{-mr\cosh(y/2)}}{\cosh(y/2)}M_{0,\nu}(y)\right].
\label{asymnear}
\end{eqnarray}
Note that, very close to the cosmic dispiration, the second term on the r.h.s is the one that dominates. The massless expression for this regime is obtained from \eqref{asymnear} by taking $m= 0$. In the case $\nu<2$ the only dominant contribution for the massive particle near the cosmic dispiration is the second one in \eqref{asymnear}, while for the massless particle the dominant contribution is also the second one, for $m=0$, but leading to the expression
\begin{eqnarray}
U_{\text{ren}}\simeq \frac{q^2\ln(2)}{8\pi^2 r},
\label{asymnearsec}
\end{eqnarray}
where the resulting $y$-integral has been solved \cite{gradshteyn2000table}. This nice expression, when the massless particle is near the cosmic dispiration, shows that the self-force is repulsive in this regime.

In the opposite limit, far away from the cosmic dispiration ($mr\gg 1$), the dominant contribution in the argument of $f_{\mu}(x)$ is $2mrs_l$ for the first term on the r.h.s of  \eqref{green4ren} and $2mrs_y$ for the second term. Note that `$m\bar{\kappa}n$' would only make this second term decrease faster, so that we can ignore it . Hence, we realize that the resulting expression is the self-energy of a massive charged scalar field in the cosmic string spacetime by noting that \cite{mota2018scalar}
\begin{eqnarray}
\sum_{n=-\infty}^{\infty}\frac{\left(n+\frac{\nu}{2}\right)}
{\left(n+\frac{\nu}{2}\right)^2+\left(\frac{y}{\phi_0}\right)^2} = \frac{\pi\sin(\nu\pi)}{\cosh(y\nu)-\cos(\pi\nu)}.
\label{SuM}
\end{eqnarray}
In other words, if the particle is placed far away from the cosmic dispiration the only dominant contribution for the self-energy would be the one due to the cosmic string. 

\begin{figure}[h]
	\centering
	{\includegraphics[width=0.7\textwidth]{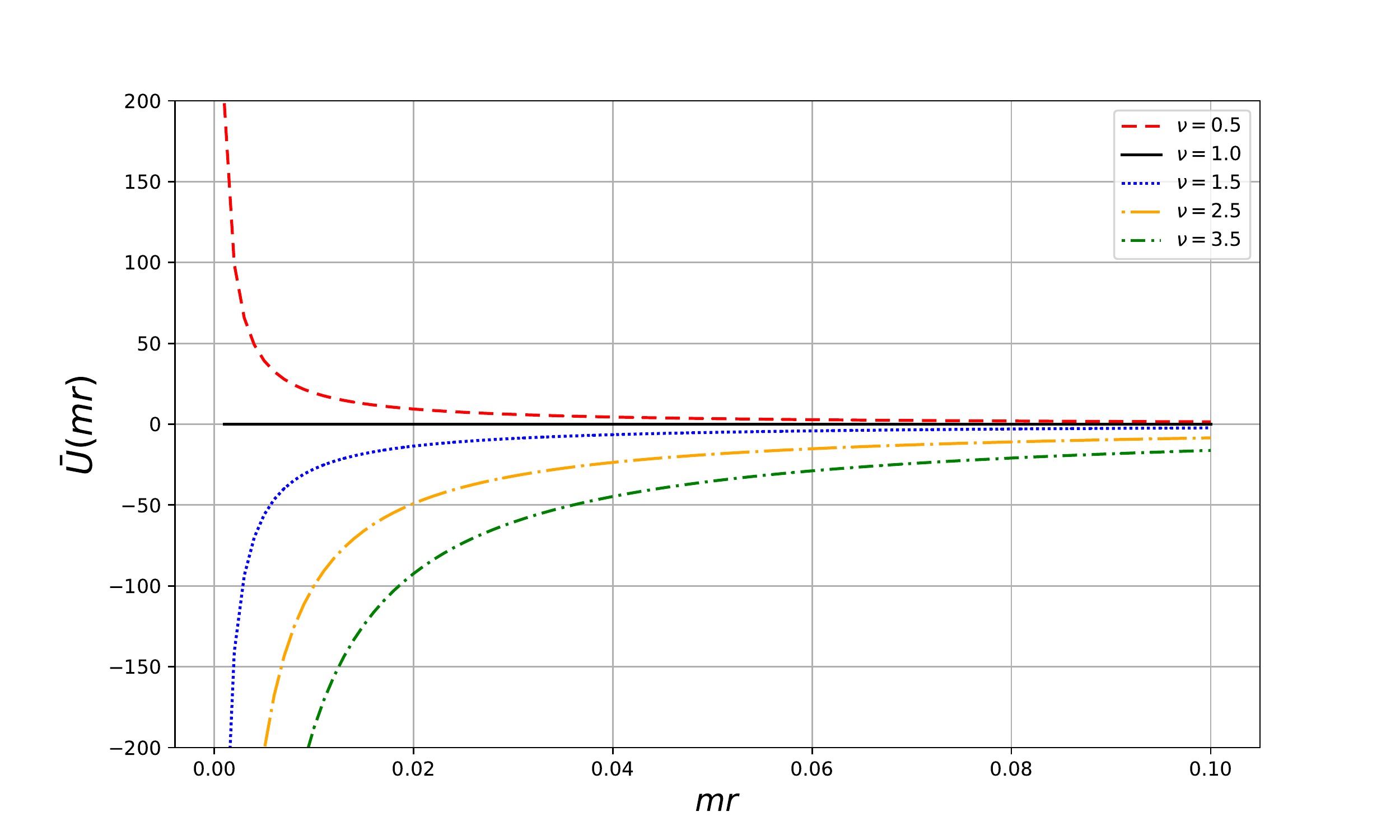}}
	\caption{Behaviour of the dimensionless massive self-energy, $\bar{U}=\frac{2(2\pi)^{\frac{3}{2}}U_{\text{ren}}}{mq^2}$, in Eq. \eqref{SEcs} as a function of $mr$.}
	\label{f3}
\end{figure}

In the absence of screw dislocation, that is, $\kappa=0$, the self-energy of a massive charged scalar field is obtained in the pure cosmic string spacetime as an exact result, which from Eqs. \eqref{green4ren} and \eqref{SuM}, is given by  
\begin{eqnarray}
U_{\text{ren}} =-\frac{mq^2}{2(2\pi)^{\frac{3}{2}}}\left[2\sideset{}{'}\sum_{l=1}^{[\nu/2]}  f_{\frac{1}{2}}\left(2mrs_l\right) - \frac{\nu\sin(\pi\nu)}{\pi}
\int_0^\infty dy \, \frac{f_{\frac{1}{2}}\left(2mrs_y\right)}{\cosh(y\nu) - \cos(\pi\nu)}\right].
\label{SEcs}
\end{eqnarray}
This general and exact expression derived here is valid for any value of the cosmic string parameter, $\nu$. For the particular values $\nu<2$, on the other hand, the first term on the r.h.s is absent. This case has been considered by Linet in Ref. \cite{Linet:1986db}. Furthermore, in the case that the cosmic string parameter assume only integer values, the second term on the r.h.s of \eqref{SEcs} vanishes, and we obtain
\begin{eqnarray}
U_{\text{ren}}=-\frac{q^2}{16\pi r}\sum_{l=1}^{\nu-1}\frac{e^{-2mr\sin(\pi l/\nu)}}{\sin(\pi l/\nu)},
\label{SEcsinteger}
\end{eqnarray}
where we have used Eq. \eqref{changesum}. Note that this contribution was not analysed in \cite{Linet:1986db}. The massive scalar self-energy in the cosmic string spacetime \eqref{SEcs} is plotted in Fig.\ref{f3} in terms of $mr$. It suggests a repulsive self-force for $\nu<1$ and an attractive one for $\nu >1$.

The massless scalar field self-energy in the cosmic string spacetime can by obtained from \eqref{SEcs} by taking the limit $m\rightarrow 0$. This provides
\begin{eqnarray}
U_{\text{ren}} =-\frac{q^2}{8\pi r}\left[\sideset{}{'}\sum_{l=1}^{[\nu/2]} \frac{1}{\sin(\pi l/\nu)}- \frac{\nu\sin(\pi\nu)}{2\pi}
\int_0^\infty dy \, \frac{[\cosh(y/2)]^{-1}}{\cosh(y\nu) - \cos(\pi\nu)}\right].
\label{SEcsmassless}
\end{eqnarray}
As the general expression above for the massless self-energy has a trivial dependence on $r$, the self-force is obtained straightforwardly. A closer inspection of Eq. \eqref{SEcsmassless} shows that the expression inside the brackets is negative for $0<\nu<1$ and positive for $\nu> 1$, resulting in repulsive and attractive self-forces, respectively. This is the opposite of the electrostatic self-force case analyzed in Refs. \cite{ escidoc:153364, linet1986force}.

For integer values of the cosmic string parameter $\nu$ we have 
\begin{eqnarray}
U_{\text{ren}}=-\frac{q^2}{16\pi r}\sum_{l=1}^{\nu-1}\frac{1}{\sin(\pi l/\nu)}.
\label{SEcsintegermassless}
\end{eqnarray}
A numerical analysis shows that the sum in $l$ above is positive for any integer value $\nu\geq 2$. This means that the massless scalar self-force is attractive, as opposed to the electrostatic case which is repulsive for these values of $\nu$, as explained above \cite{ escidoc:153364, linet1986force}. 

It is notable that even when a point particle, in our case scalar, is set at rest it feels the influence of its own created classical field, something that does not happen in Minkowski spacetime. This is a consequence of the particle be considered in a curved spacetime, as showed here in the case of a cosmic dispiration spacetime. The results for the self-energies of massive \eqref{green4ren} and massless \eqref{green4renmassless} charged scalar point particles, including the the case $\nu=1$ (only screw dislocation), have not been previously obtained in literature. Note that the authors in Ref. \cite{DeLorenci:2001sn} analyzed only the cases of electrostatic and gravitational self-energies. The case of a pure cosmic string, that is, $\kappa=0$, considering a massive scalar point particle, has been considered in \cite{Linet:1986db} only for $\nu<2$. Here, we have shown to exist a contribution when $\nu>2$, given by the first term on the r.h.s of \eqref{SEcs}. Hence, when the cosmic string parameter $\nu$ is assumed to be given by integer numbers, the only contribution to the self-energies come from \eqref{SEcsinteger} and \eqref{SEcsintegermassless}. 
\section{Conclusion}
\label{Con}
We have investigated the classical self-interaction associated with a minimally coupled charged scalar point particle placed at rest in the cosmic dispiration spacetime, also known as twisted cosmic string. In order to solve the equation of motion sourced by the charged scalar particle we have obtained an exact expression for the Green's function valid for any value of both the cosmic string and screw dislocation parameters. We then identified the divergent term of the Green's function, subtracted it out and obtained the renormalized self-energy. Important features associated with the nature of the self-interaction of the point particle, in the cosmic dispiration spacetime, have been shown in the plots presented in Figs.\ref{f1} and \ref{f2} for the massive and massless cases, respectively. To the best of our knowledge, these results have been first obtained in the present paper. 

We have also analyzed the limiting case in which there is only a screw dislocation, that is, $\nu=1$, as well as the limiting case in which there is only the cosmic string, i.e., $\kappa=0$. In particular, in the cosmic string spacetime, we compared our results with the one obtained by Linet in \cite{Linet:1986db} and pointed out a new contribution obtained here, coming from the values $\nu\geq 2$. The general expression for the massive self-energy in the cosmic string spacetime has been plotted in Fig.\ref{f3}. In the massless case, we have verified that for $0<\nu<1$ and for $\nu> 1$, the resulting self-forces are repulsive and attractive, respectively. This is the opposite of what happens in the electrostatic case, considered in Refs. \cite{ escidoc:153364, linet1986force}. Moreover, the self-force of the scalar point particle in the screw dislocation spacetime can be inferred from the left plot in Fig.\ref{f1} to be repulsive for $m\kappa=0.25$. The plot in Fig.\ref{f2} also seems to indicate the same in the massless case.

\section*{Acknowledgments}
K.E.L.F would like to thank the Brazilian agency CAPES for financial support. E.A.F.B is grateful for the hospitality of the Departament of Physics of Universidade Federal de Pernambuco, where part of this work was done. E.A.F.B also thanks the
Brazilian agency CAPES for financial support. H.F.S.M acknowledges partial support from the Brazilian agency CNPq (Grants No. 305379/2017-8 and 430002/2018-1).


\end{document}